\documentclass[
 reprint,
 amsmath,amssymb,
 aps,
superscriptaddress,
]{revtex4-1}

\usepackage{graphicx}
\usepackage{dcolumn}
\usepackage{bm}
\usepackage{hyperref}

\begin{document}

\title{Slow light-enhanced optical imaging of microfiber radius variations with sub-\AA ngstr\"om precision}

\author{Michael Scheucher}
\altaffiliation{These authors contributed equally}
\affiliation{TU Wien, Atominstitut, Stadionallee 2, 1020 Vienna, Austria}
\author{Khaled Kassem}
\altaffiliation{These authors contributed equally}
\affiliation{TU Wien, Atominstitut, Stadionallee 2, 1020 Vienna, Austria}
\author{Arno Rauschenbeutel}
\affiliation{TU Wien, Atominstitut, Stadionallee 2, 1020 Vienna, Austria}
\affiliation{Department of Physics, Humboldt-Universit\"at zu Berlin, 10099 Berlin, Germany}
\author{ Philipp Schneeweiss}
\affiliation{TU Wien, Atominstitut, Stadionallee 2, 1020 Vienna, Austria}
\affiliation{Department of Physics, Humboldt-Universit\"at zu Berlin, 10099 Berlin, Germany}
\author{J\"urgen Volz}
\email{juergen.volz@hu-berlin.de}
\affiliation{TU Wien, Atominstitut, Stadionallee 2, 1020 Vienna, Austria}
\affiliation{Department of Physics, Humboldt-Universit\"at zu Berlin, 10099 Berlin, Germany}

\date{\today}

\begin{abstract}
Optical fibers play a key role in many different fields of science and technology. In particular, fibers with a diameter of several micrometers are intensively used in photonics. For these applications, it is often important to precisely know and control the fiber radius. Here, we demonstrate a novel technique to determine the local radius variation of a 30-micrometer diameter silica fiber with sub-\AA ngstr\"om precision with axial resolution of several tens of micrometers over a fiber length of more than half a millimeter. Our method relies on taking an image of the fiber's whispering-gallery modes (WGMs). In these WGMs, the speed of light propagating along the fiber axis is strongly reduced. This enables us to determine the fiber radius with a significantly enhanced precision, far beyond the diffraction limit. By exciting different axial modes, we verify the precision and reproducibility of our method and demonstrate that we can achieve a precision better than 0.3 \AA. The method can be generalized to other experimental situations where slow light occurs and, thus, has a large range of potential applications in the realm of precision metrology and optical sensing.
\end{abstract}

\pacs{Valid PACS appear here}
\maketitle
\section{Introduction}
The resolution of optical far-field imaging systems is typically limited by diffraction. As a consequence, it is not possible to resolve details with a spacing smaller than half the wavelength of the imaging light. Even for high-end objectives with a numerical aperture $N\!A\approx 1$, this limits the resolution to a few hundred nanometer when using visible light. At the same time, the position of point-like scatterers or emitters can be determined with almost arbitrary precision by fitting the point-spread function of the imaging system to the image \cite{Thompson2002}. This technique is, e.g., used in super-resolution microscopy \cite{Hell2007} or quantum technology \cite{Wong2016} and yields a position accuracy down to a few nanometers. Other super-resolution techniques rely on the spatially selective deactivation of fluorophores and are thus restricted to be used in combination with certain emitters \cite{Hell2007}. Despite of these advances, the observation of sub-wavelength structures and size variations of macro- and microscopic objects is still challenging in many settings.\\
For example, most fabrication processes of optical fibers do not allow one to directly monitor the local fiber radius. Nevertheless, its precise knowledge is important for many applications.
A standard method to determine the size of objects with a resolution at the nanometer scale is scanning electron microscopy (SEM). However, this technique requires inserting the fibers into a vacuum recipient which limits the cycle time for analyzing fibers. Several methods that use fiber-guided light to determine the fiber radius have been demonstrated \cite{Wiedemann2010,Sumetsky2011,Hoffman2015,Semenova2015,Keloth2015,Madsen2016}. However, none of these methods provides simultaneously good axial and nanometer-scale radial precision. Scanning methods using external probes, via e.g., a second fiber, enable good axial and radial resolution \cite{Birks2000,Sumetsky2010a}. However, they fall short of providing means to characterize many samples in a short time as the probe fiber has to be mechanically moved along the sample. Furthermore, for those approaches, the probe and the sample fiber are in direct mechanical contact which potentially damages the sample when the two fibers are sliding over each other or when separating the fibers again. Other approaches, including measuring the force--elongation curve \cite{Holleis2014} or optical imaging methods \cite{VanderMark1994,Warken2004}, do not achieve sufficient precision.\\
Here, we demonstrate a novel method that employs slow light for imaging an optical fiber. The method allows us to determine the radius profile of a sample microfiber with sub-\AA ngstr\"om precision while providing a high axial resolution at the same time.
Slow light has attracted significant interest in  recent years \cite{Krauss2008}, resulting in various applications.
For example, slow light emerging in photonic crystals has been utilized for realizing sensors \cite{Shi2007,Qin2016,Kraeh2018}, amplifiers \cite{Ek2014}, enhancement of optical power densities \cite{McMillan2006,McGarvey-Lechable2014,Yan2017}, and nonlinear optics  \cite{Corcoran2009,Xiong2011}.
\\
In this work, we make use of slow light arising in the presence of whispering-gallery modes (WGMs) which naturally form around optical fibers \cite{Sumetsky2013}. WGMs are optical resonances, in which light travels along the circumference of the fiber, undergoing continuous total internal reflection \cite{Matsko2006}. Due to the mostly azimuthal light propagation, the effective group velocity of light in axial direction along the fiber is significantly reduced. In this situation, small radius variations already give rise to a strong axial potential for the light and, thus, determine the axial structure of the WGMs (see Fig.~\ref{fig:Dispersion_relation}a). The slow group velocity in axial direction results in a small axial wavevector. When the light forms a standing wave, this gives rise to intensity modulations of comparably large spatial period of the wavefunction. By imaging the light tunneling from these axial standing-wave WGMs into free-space using a CCD camera, we measure the axial mode profile from which we can precisely determine the radius profile of an optical fiber over a long segment of the fiber in single shot operation.

\section{Theoretical model}

\begin{figure}[tb]
	\centering
		\includegraphics[width=0.5	\textwidth]{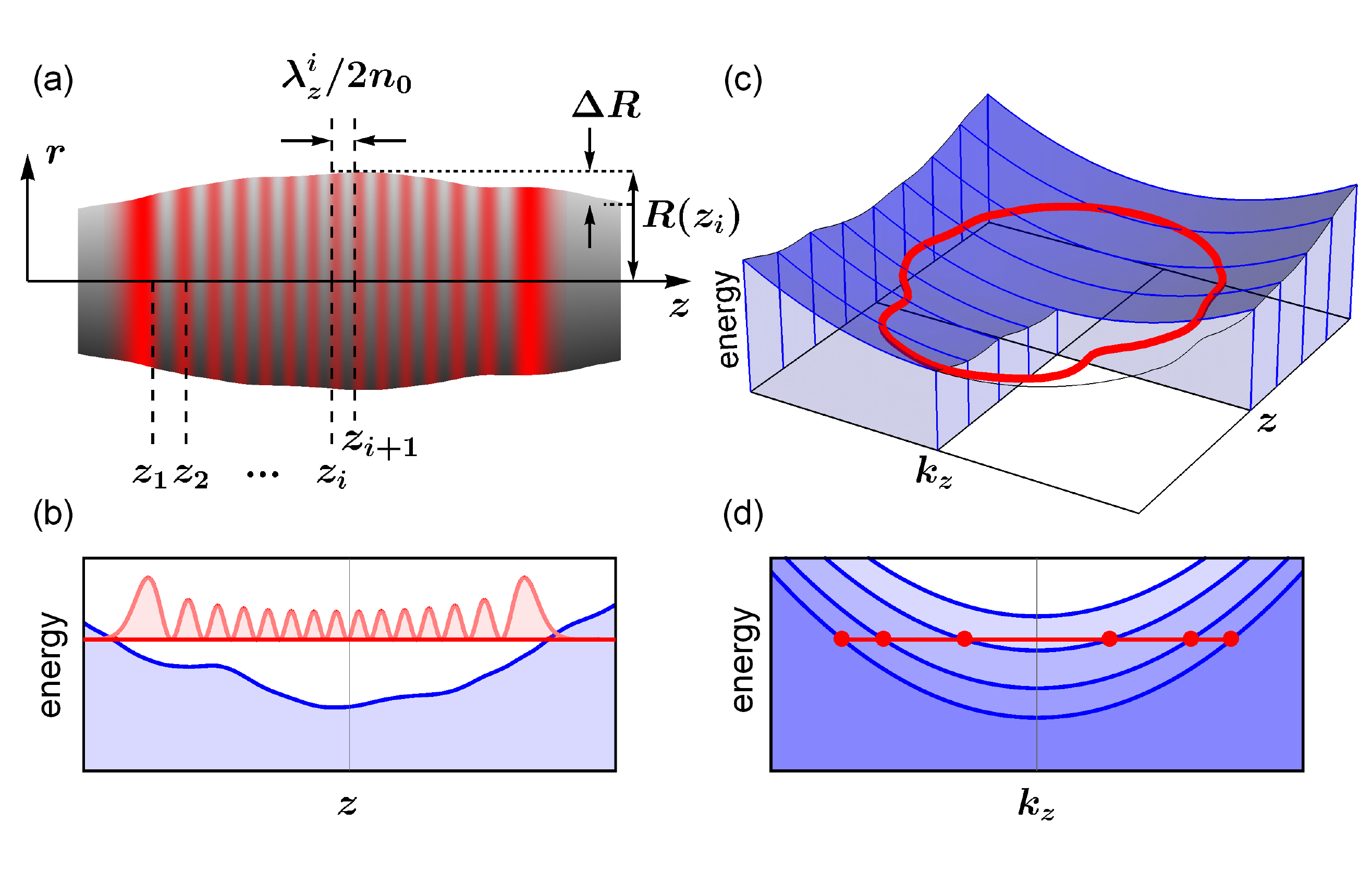}
\caption{(a) Side view of a fiber with nanoscale radius variation along the fiber axis (the radius variations are strongly exaggerated for illustration purposes). Due to these variations, a characteristic mode structure forms with clear intensity maxima and minima. The distance between two adjacent intensity minima is given by the axial wavelength $\lambda_z^i/2n$. For the sample fiber considered here, the central radius $R_0$ is a few tens of micrometer and the radius variation $\Delta$r is on the order of a few nanometer. (b) The radius profile can be translated into a potential energy $V(z)$, which can support optical bound states, i.e. localized WGMs, indicated in red. (d) Axial dispersion relation of the photon energy $\hbar \omega(k_z)$ for different fiber radii. (c) Phase space representation of a WGM with energy $\hbar \omega_q$ that is trapped in the potential formed by nanoscale radius variations along $z$. Projecting the plot onto the $z$ and $k_z$ axis yields the cases shown in (b) and (d), respectively.}
	\label{fig:Dispersion_relation}
\end{figure}

When light circulates around the circumference of a fiber, the electric field $ \boldsymbol{\mathcal{E}}$ has to fulfill the Helmholtz equation
\begin{equation}
(\nabla ^2 + k^2n^2(r)) \boldsymbol{\mathcal{E}}=0\;.
\end{equation}
Here, $k=2\pi/\lambda$ is the light's wave number and $n(r)$ describes the refractive index inside ($n=n_0$) and outside ($n\approx 1$) of the fiber. When the radius variations of the fiber are sufficiently small, we can separate the propagation of the light into a propagation parallel to the fiber axis with wavevector $\boldsymbol{k}_z$ and along the fiber's circumference with wavevector $\boldsymbol{k}_{\phi,r}$, where $\boldsymbol{k}=\boldsymbol{k}_{\phi,r}+\boldsymbol{k}_z$. This allows us to separate the Helmholtz equation using the Ansatz $\boldsymbol{\mathcal{E}}=\Phi(\phi,r)\mathcal{Z}(z)$. For a given radius $R(z)$, the radial and azimuthal equations can locally be solved, yielding the resonance condition $k_{\phi,r}(z)=f(m,p) m/(R(z) n_0)$, where $m$ and $p$ are the azimuthal and radial quantum numbers of the WGM, respectively, and $f(m,p)\sim1$ is a factor that describes the geometrical dispersion of the fiber, see supplemental material. The axial part of the system can be treated as a one-dimensional (1D) problem, described by the axial wave equation
\begin{equation}
\left(\partial_z^2 + k^2n^2-f^2\frac{m^2}{R(z)^2}\right) \mathcal{Z}(z)=0\;,
\label{eq:axial_wave_equation}
\end{equation}
where we made use of $k^2=k_z^2+k_{\phi,r}^2$. Equation~(\ref{eq:axial_wave_equation}) is formally identical to a 1D Schr\"odinger equation \cite{Sumetsky2011} and describes the propagation of a photon with effective mass $m_{eff}=\hbar kn_0^2/2c$ in the potential landscape that is set by the radius profile $R(z)$ with the potential $V(z)/(\hbar^2/2m_{eff})=(m\,f)^2/R(z)^2$ (see Fig.~\ref{fig:Dispersion_relation}b and  Fig.~\ref{fig:Dispersion_relation}c), where $c$ is the vacuum speed of light. The local group velocity $c_{gr}$ of light traveling along this potential can be obtained from the dispersion relation $\omega(k_z)=c/n_0\cdot\sqrt{k_{\phi,r}^2+k_z^2}$ (see Fig.~\ref{fig:Dispersion_relation}d) and is approximated to be  $c_{gr}=d\omega/dk_z\approx c/n_0 \cdot k_z/k$  \cite{Sumetsky2013}. For light fields close to the band edge $k\approx k_\phi$, one obtains a very strong reduction in group velocity, i.e. $c_{gr}\ll c$. Due to this group velocity reduction, the phase velocity of the light in axial direction $c_{ph}=c^2/(n_0^2c_{gr})$ is strongly increased. As the axial wavelength, $\lambda_z$, is directly proportional to $c_{ph}$, $\lambda_z$ is significantly larger than the vacuum wavelength. Due to this magnification effect, the axial wavelength can now be measured with high accuracy, which in turn allows one to perform a high-accuracy measurement of the axial potential of the light and, thus, of the corresponding fiber radius variations. The dependency of the local group velocity and, thus, of $\lambda_z$ on the fiber radius can be expressed as
\begin{equation}
R(z)=f(m,p)\frac{m \lambda}{2\pi\, n_0} \left(1-\mathcal{S}(z)^2\right)^{-\frac{1}{2}}\;, \label{eq:radius}
\end{equation}
where we introduced the velocity reduction $\mathcal{S}=c_{gr}/c=\lambda/\lambda_z$. In our experiment, the axial radius variations that define the potential $V(z)$ result in bound states in axial direction. For this case, the light oscillates between the two axial turning points, called the caustics, creating a standing wave along the fiber, while still propagating along the circumference as a running wave. The number of nodes of the standing wave are labeled with the quantum number $q$. Measuring the intensity distribution of the axial standing wave allows one to determine the local axial wavelength $\lambda_z$ (see Fig.~\ref{fig:Dispersion_relation}a) and, thus, the local fiber radius with enhanced precision. The enhancement factor can be expressed as 
\begin{equation}
M=\left|\frac{\partial \lambda_z}{\partial R(z)}\right|\approx\
\frac{ 2\pi}{m \,f}\frac{1}{\mathcal{S}^3}\;.
\label{eq:mag}
\end{equation}
From Eq.~(\ref{eq:mag}), it follows that the enhancement factor and, thus, the possible precision, strongly increases with decreasing $\mathcal{S}$ (increasing $\lambda_z$). The radius variation can be determined, at best, with precision
\begin{equation}
\Delta R =\frac{\Delta \lambda_z}{M}\;.
\label{eq:resolution}
\end{equation}
where $\Delta \lambda_z$ is the error in determining $\lambda_z$ which is given by the optical resolution of the imaging system. Equation~(\ref{eq:resolution}) can be seen as a direct consequence of the uncertainty principle which prevents one from measuring both, position and momentum,
with arbitrary precision. For typical experimental parameters, values of $M=10^4...10^5$ can easily be achieved, see Table \ref{tab:MeasuredAxialModes}. We note that the enhanced precision of the measurement of the radius variation given in Eq.~(\ref{eq:resolution}) is inherently connected with a decrease in axial resolution which is approximately given by $\lambda_z/2n_0$. 

\section{Setup and measurement procedure}

\begin{figure}[tp]
	\includegraphics[width=0.47\textwidth]{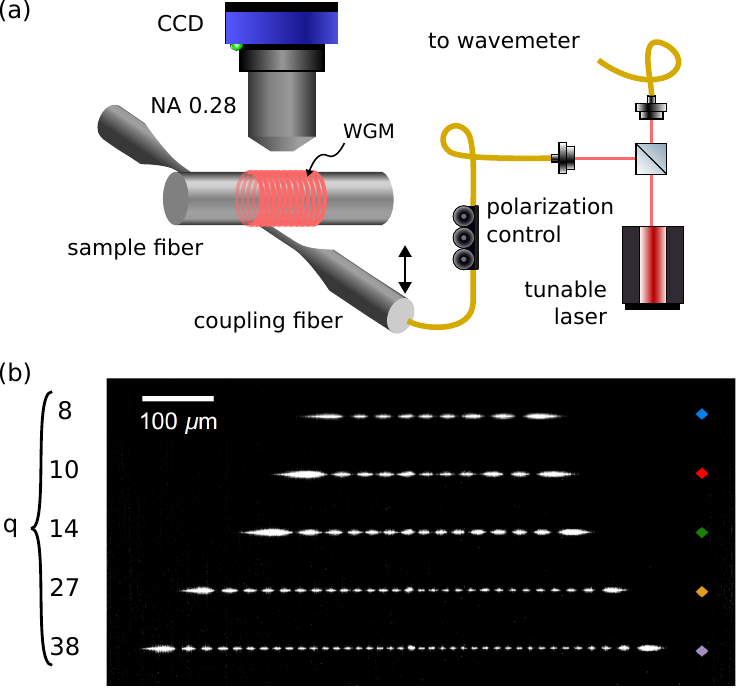}
	\caption{(a) The experimental setup used to determine the radius profile of the sample fiber consists of a tunable diode laser, polarization controllers, a coupling fiber, and a CCD camera with an NA=0.28 objective. For further diagnostics, an additional photo diode can be used to monitor the fiber transmission, and the incident light is sent onto a wavelength meter. (b)~Micrographs of light tunneling out of the WGMs in direction of the camera for different axial modes, $q~\in~\{8,10,14,27,38\}$, which are used for determining the local fiber radius in Fig.~\ref{fig:results} }
	\label{fig:schematic}
\end{figure}

In the following, the fiber under examination is referred to as sample fiber. In order to couple light into the sample fiber, we use a tapered fiber coupler. This coupling fiber is mounted on a translation stage and aligned perpendicular to the sample fiber. This allows us to evanescently couple light from the coupling fiber to the  sample fiber, see Fig.~\ref{fig:schematic}a. The coupling rate between the fibers can be adjusted via their relative distance. 
When light is sent through the coupling fiber and its frequency is scanned, WGMs are excited whenever the resonance condition is fulfilled and when the WGM has a finite mode overlap with the evanescent field of the coupling fiber. 
The excited WGMs can be observed using a CCD camera that images the mode structure along the sample fiber, as shown in Fig.~\ref{fig:schematic}b. We emphasize that the collected light is not scattered from the surface, but originates from tunneling of WGMs through the potential barrier imposed by the refractive index jump at the fiber surface \cite{Tomes2009}. Hence, the imaging of the mode structure does not rely on local surface pollution or roughness but occurs for any structure supporting WGMs. For imaging, we employ a CCD camera (mvBlueFox3, Matrix Vision) in combination with a standard microscope objective (Mitutoyo 10X M Plan APO LWD) with NA=0.28. The camera system and the coupling fiber are mounted on opposite sides of the sample fiber (see Fig.~\ref{fig:schematic}a). This ensures a clear view of the WGMs using the camera. For the excitation of the WGMs, we used a tunable diode laser (Velocity Laser 6316, New Focus) with a wavelength of about 845 nm. \\
In order to infer the radius profile from the camera images such as shown in Fig.~\ref{fig:schematic}b, we use the following procedure. We determine the axial intensity profile of the WGM by averaging over several horizontal pixel lines. Then the positions of zero intensity $z_i$ are extracted from a parabolic fit to the data points in the vicinity of the intensity minima. For our imaging system, this procedure enables us to determine these positions with a precision of $\Delta \lambda_z= 0.2$ $\mu$m, see supplemental material. In order to obtain the absolute value of the local radius, according to Eq.~(\ref{eq:radius}), we also require the azimuthal and radial quantum numbers $m$ and $p$ of the imaged WGM. Therefore, we compare the measured azimuthal free spectral range of different mode families to the free spectral range obtained from numerically solving the radial wave equation of a dielectric cylinder, see supplemental material.

\section{Results and precision}
\begin{figure}
	\centering
		\includegraphics[width=\linewidth]{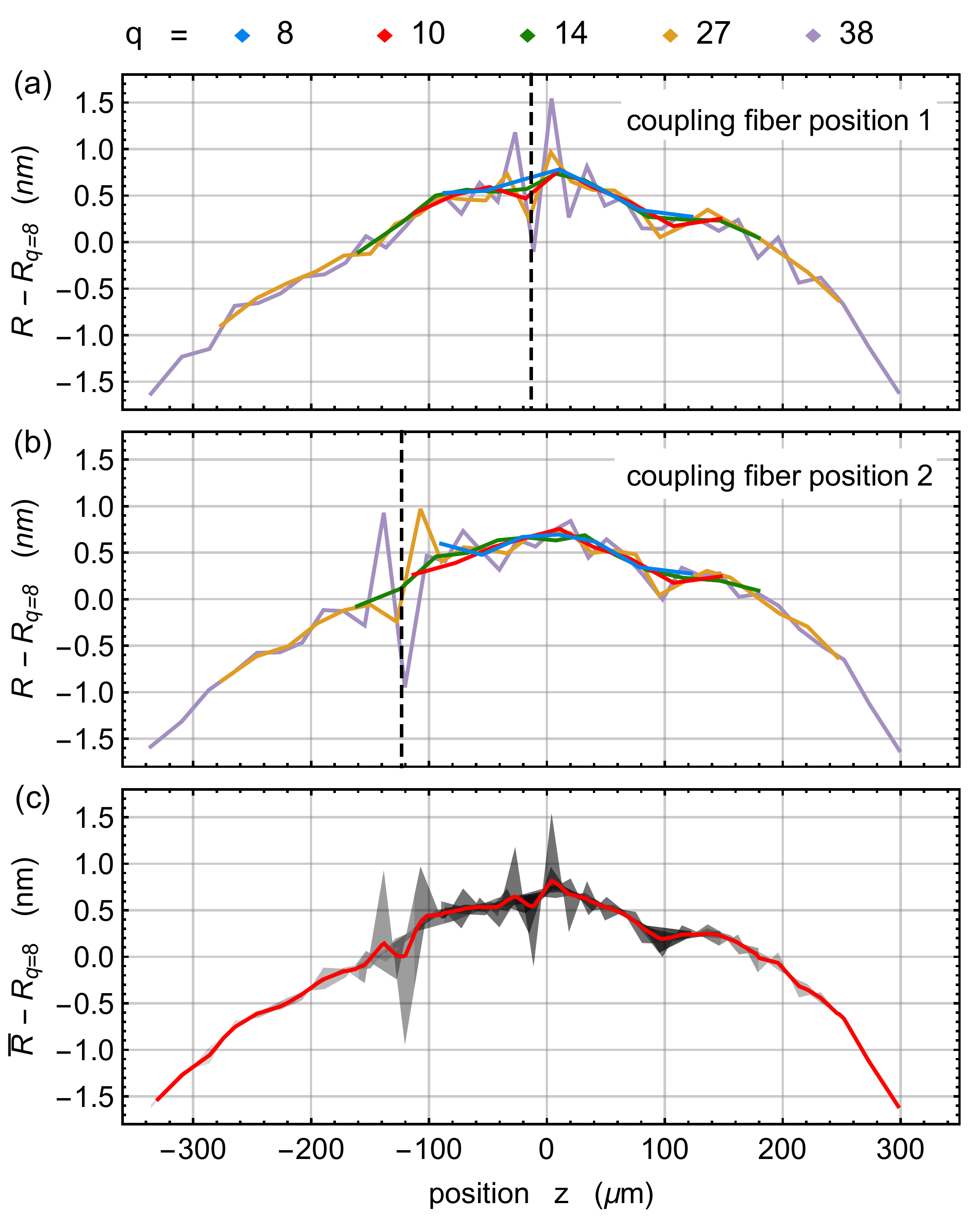}
	\caption{(a) Evaluated radius profiles as a function of the position along the sample fiber for WGMs with different axial quantum number $q$. The position of the coupling fiber is indicated by the dashed vertical line. (b) Same as (a) only with the coupling fiber moved by $\sim$110~$\mu$m. (c) Mean radius profile obtained when all measurements are taken into account (red line). The shaded area indicates the deviation of the individual measurements from the mean profile. For better visibility, we subtracted the (constant) caustic radius of the sample fiber for the $q=8$ mode, $R_{q=8}=15.475(1)$~$\mu$m.}
	\label{fig:results}
\end{figure}
For a sample fiber of approximately 30~$\mu$m  diameter, we recorded two sets of images with the CCD camera. First, the coupling fiber is placed approximately at the center of the sample fiber waist, and we perform five independent measurements of the radius profile by exciting five TE-polarized modes with different axial quantum number $q$ but the same quantum numbers $m$ and $p$. Figure~\ref{fig:schematic}b shows micrographs of the light tunneling out of the fiber for these WGMs. The resonance wavelengths, measured with the wavelength meter (High Finesse WS7-60), as well as the corresponding $q$ are summarized in Tab.~\ref{tab:MeasuredAxialModes}. The azimuthal and radial quantum numbers have been determined to be $m=117$ and $p=8$, which corresponds to a factor $f(8,117)=1.42605843(1)$, see supplemental material. The radius profiles of the fiber extracted from these images are shown in Fig.~\ref{fig:results}. The radius profiles obtained using different modes are in very good agreement with each other, illustrating the reproducibility of the method. It also becomes evident that with increasing axial resolution (for higher $q$), the radial resolution decreases, as expected for this measurement (see Eq.~(\ref{eq:resolution})). In order to check if the position of the coupling fiber alters the measured radius profile, the measurement was repeated after moving it by $\sim 110$~$\mu$m in axial direction. The results are shown in Fig.~\ref{fig:results}b and are in very good agreement with the  measurements in Fig.~\ref{fig:results}a. In the measurements with the highest axial resolution, i.e., with the highest values of $q$, one observes a small distortion at the position of the coupling fiber. However, the errors introduced by scattered light from the coupling fiber are well below a nanometer and can be corrected by performing two measurements with different fiber positions. It should be noted that the obtained radius is an azimuthal average and, thus, the measurement does not provide information about the ellipticity of the fiber cross section.

\begin{table}[htbp]
\centering
\caption{\bf Axial quantum number $q$, the resonance wavelength $\lambda_{vac}$, mean axial wavelength $\overline{\lambda_z}/n_0$, and the resulting mean enhancement factor $\overline{M}$ of the WGMs under examination. }
\begin{tabular}{ccccc}
\\
\hline
\hspace{0.5cm}$q$\hspace{0.5cm} & $\lambda_{vac}$ (nm) & $\overline{\lambda_z}/n_0$ ($\mu$m)&$\overline{\mathcal{S}}$ & $\overline{M}\, /\, 10^3$ \\ \hline
 8 & 846.4819 & 68.8 	& 0.0058	&190 \\
 10 & 846.4724 & 63.5 & 0.0063	&149 \\
 14 & 846.4569 & 55.8 & 0.0072	&101 \\
 27 & 846.4007 & 40.9 & 0.0098	&40 \\
 38 & 846.3500 & 34.4 & 0.0117	& 24 \\
\hline
\end{tabular}
  \label{tab:MeasuredAxialModes}
\end{table}

Two different types of errors occur when measuring the radius profiles $R(z)$. On the one hand, there is a systematic error in the absolute radius $R_q=fm\lambda_q/n_0$ which amounts to an uncertainty of around $\pm 1$~nm in our case. This error is identical for each measurement and originates from the systematic error in determining $\lambda$, $f$ and $n_0$ in Eq.~(\ref{eq:radius}). For our setup, this is dominated by the limited knowledge of the refractive index of the sample fiber, $n_0$. On the other hand, the radius variations along the fiber, $R(z)-R_q$, can be measured with much higher accuracy and are, in our measurement, limited by the measurement accuracy of the axial wavelength $\lambda_z$, see supplemental material for more details.

\begin{figure}
	\centering
		\includegraphics[width=\linewidth]{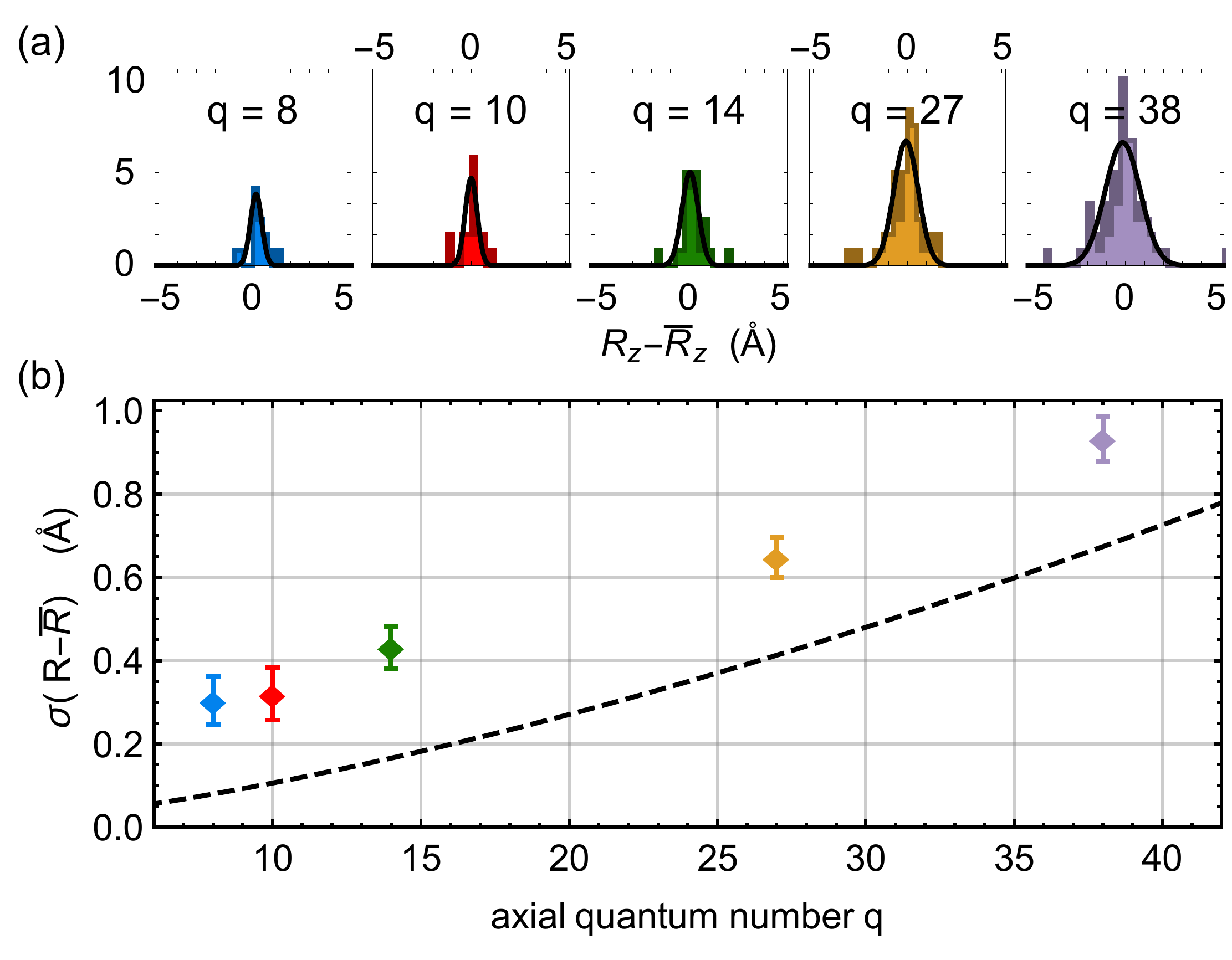}
	\caption{(a) Histogram of the deviation for each individual measurement from the average radius profile $\overline{R}$ for different $q$.  (b) Standard deviation of $R-\overline{R}$, obtained from the Gaussian fits in (a), as a function of $q$. The error bars show the standard errors for the parameter estimates of the fit. The dashed line corresponds to the theoretically expected mean standard error calculated assuming a parabolic radius profile, which increases with rising $q$. }
	\label{fig:discussion}
\end{figure}

From, in total, 10 individual measurements, we determine the most likely fiber radius profile $\bar{R}$ by linearly interpolating the measurement points for each mode and taking the average over all modes. The resulting profile is shown in Fig.~\ref{fig:results}c. It indicates an almost perfect cylinder that has a residual radius variation of $~$2~nm over an axial extend of 600~$\mu$m. \\
In order to get an estimate of the precision of the individual measurements, we compute the deviation between each measurement point from the most likely radius profile,  $R-\bar{R}$, and plot a histogram of the deviations for each axial mode, see Fig.~\ref{fig:discussion}a. The standard deviation of this difference  $\sigma(R-\bar{R})$ gives an estimate of the error of our measurement of the radius variation along the fiber. Figure~\ref{fig:discussion}b shows this standard deviation as a function of $q$ and a theoretical prediction for comparison (dashed line). For calculating the theory curve, we approximate the radius profile by a parabolic profile from which we then estimate the average axial wavelength. Taking into account all known experimental errors, we derive the expected mean standard deviation of the axial radius profile, see supplementary material.
Our analysis shows that the measurement precision of the radius variation increases with decreasing $q$, as expected. For all modes, we observe sub-\AA ngstr\"om precision, and for $q=8$ we reach a precision of $0.30\pm0.06$ \AA. Our precision inferred from the measurement data shows the same trend with $q$ as the theoretical prediction.

\section{Summary and Outlook}
In summary, we demonstrated a method that uses slow light to determine the radius variations along an optical fiber with sub-\AA ngstr\"om precision. In contrast to other fiber-probing methods, the demonstrated approach does only require approximate knowledge of the system parameters, e.g., of the quantum numbers of the excited optical modes utilized for imaging. As discussed in the supplementary material, it can be performed using low cost equipment, such as a standard microscope objective and a basic CCD camera, while still maintaining sub-\AA ngstr\"om precision. Furthermore, this method minimizes systematic errors and possible damage by evanescently interfacing the sample fiber. In addition, we are also able to measure the absolute value of the fiber radius, where we exceed the accuracy of most other methods.\\
Importantly, our results apply to any fiber or in general any optical system that supports WGMs, as long as a partial standing wave is formed along the fiber that can be resolved within the dynamic range of the camera.
Furthermore, it can be generalized to many situations where slow light occurs. Thus, our low-cost and non-destructive approach might be of use for technical applications such as in-situ monitoring of fiber and microfiber fabrication. Finally, it could be utilized for spatially-resolved sensing \cite{Foreman2015}.

\section*{Acknowledgments}

The authors are grateful to T. Hoinkes for technical support.
This work has received funding from the European Commission under the projects ErBeStA (No. 800942) and ERC grant NanoQuaNt, the Austrian Academy of Sciences (\"OAW, ESQ Discovery Grant QuantSurf) as well as the Austrian Science Fund under the project NanoFiRe (No. P 31115). 
\\

\bibliography{bib}
\end{document}


\title{Supplemental material}

\maketitle
\section{Geometric dispersion}

\begin{figure}[h!]
\centering
\includegraphics[width=\linewidth]{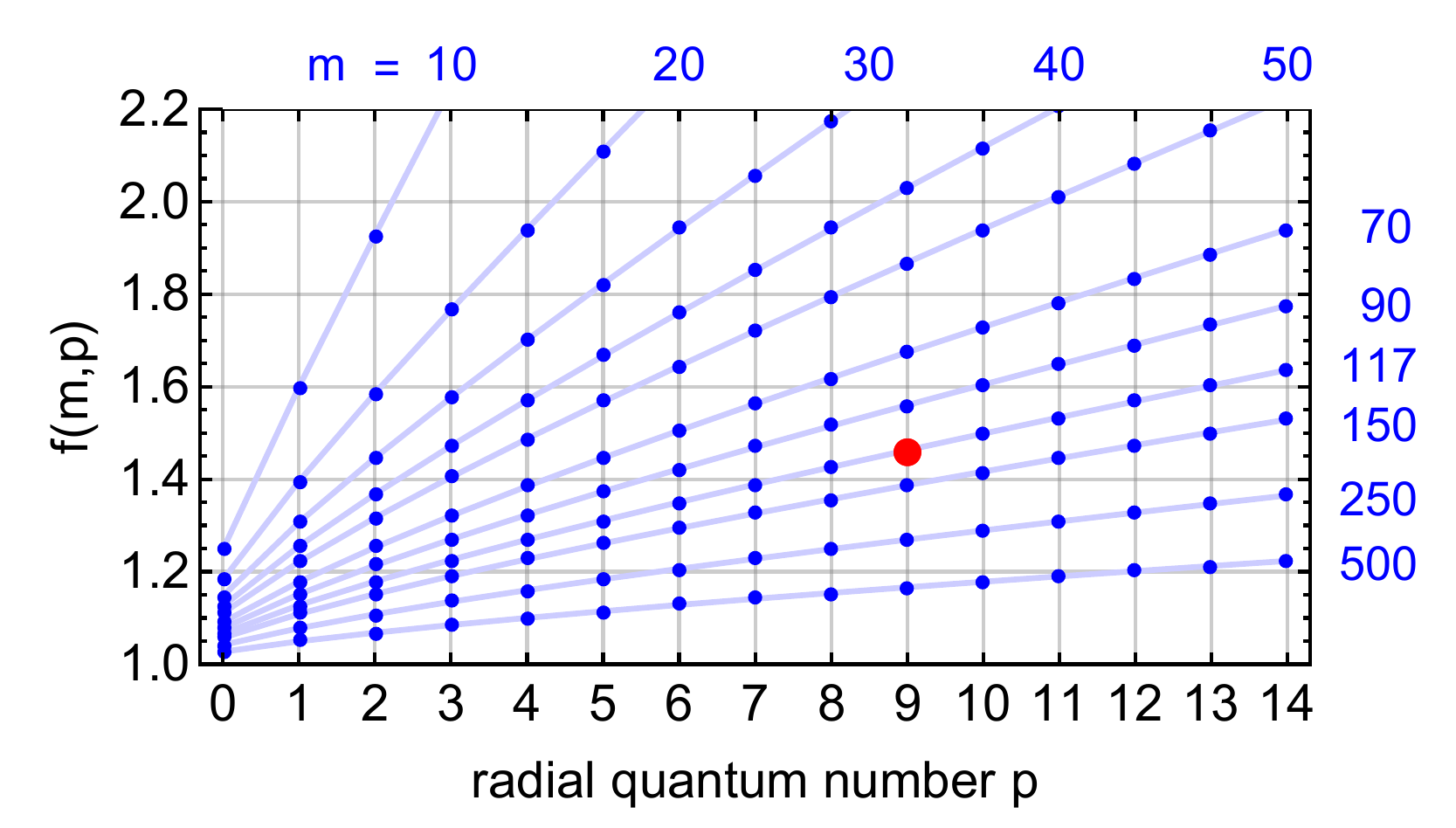}
\caption{Geometric dispersion factor $f(m,p)$ for different azimuthal quantum numbers $m$ as a function of the radial quantum number $p$ for a refractive index $n_0=1.4526$. The red dot indicates the value for the modes used in this manuscript.}
\label{fig:geom_disp}
\end{figure}

The dispersion of light in whispering-gallery modes (WGMs) depends not only on the resonator's material but also on its geometry. This stems from the fact that the light's radial distribution, inside and outside of the resonator material, for a given wavelength, is strongly related to the resonator's geometry. As a consequence, when changing the resonators dimensions the resonance spectrum also changes. In the following, the geometric dispersion will be treated by solving the radial wave equation.\\
By making use of the cylindrical symmetry and separating the Helmholtz equation Eq.~(1), the radial differential equation, after solving the azimuthal part, is 
\begin{equation}
\partial^2_r \Phi(r) +\frac{1}{r}\partial_r \Phi(\phi,r)+\left(k^2_{\phi,r}-\frac{m^2}{r^2}\right)\Phi(\phi,r)=0\;.
\end{equation}
Here, $(r, \phi,z)$ are the cylindrical coordinates, $\Phi(r)$ is the radial wave function, $m$ the azimuthal quantum number, and  $k_{\phi,r}$ the component of the wave vector in the $(\phi,r)$-plane.
The solutions are given by the Bessel function of the first kind, $J_m$, and the Bessel function of the second kind, $Y_m$.
Resonances of a cylindrical resonator with radius $R$ can then be found by solving a transcendental equation, that implements the  boundary conditions imposed by Maxwell's equations. The transcendental equation reads \cite{Oraevsky2002}
\begin{equation}
P\frac{J'_m(kn_0R)}{J_m(kn_0R)}=\frac{Y'_m(kR)}{Y_m(kR)}\;,
\label{eq:transcendental}
\end{equation}
where $P=n_0$ or $1/n_0$ for TE and TM modes, respectively and $n_0$ is the refractive index of the resonator.
An exact analytic solution of Eq.~(\ref{eq:transcendental}) is generally not possible, such that resonance frequencies must be determined either numerically or by means of analytical approximations.
The resulting component of the wave vector can be expressed as 
\begin{equation}
k_{\phi,r}(z) =f(m,p)\frac{m}{n_0\, R(z)}\;,
\label{eq:kphi}
\end{equation}
where $f(m,p)$ accounts for the geometric dispersion, and the radial quantum number, $p$, indicates the $p$th root of Eq.~(\ref{eq:transcendental}). In order to illustrate the dependencies of $ f(m,p)$, it can be approximated by, e.g., the asymptotic expansion \cite{Lam1992} 
\begin{align}
f(m,p)\approx&\,1+\frac{\alpha_p }{2^{1/3}m^{2/3}}-\frac{P}{m (n_0^2-1)^{1/2}}+\frac{3}{10}\frac{\alpha_p^2}{2^{2/3} m^{4/3}}\nonumber\\&
-\frac{P(n_0^2-2P/3)}{(n_0^2-1)^{3/2}}\frac{\alpha_p}{2^{1/3}m^{5/3}}+ ...\;,
\label{eq:Lam}
\end{align}
where $\alpha_p$ are the $p$th root of the Airy function \cite{Abramowitz1964}. In Figure~\ref{fig:geom_disp}, the geometric dispersion factor $f(m,p)$ is plotted for different quantum numbers $m$ and $p$.

\section{Determination of the radial and azimuthal quantum number}
\begin{figure}[htbp]
\centering
\includegraphics[width=\linewidth]{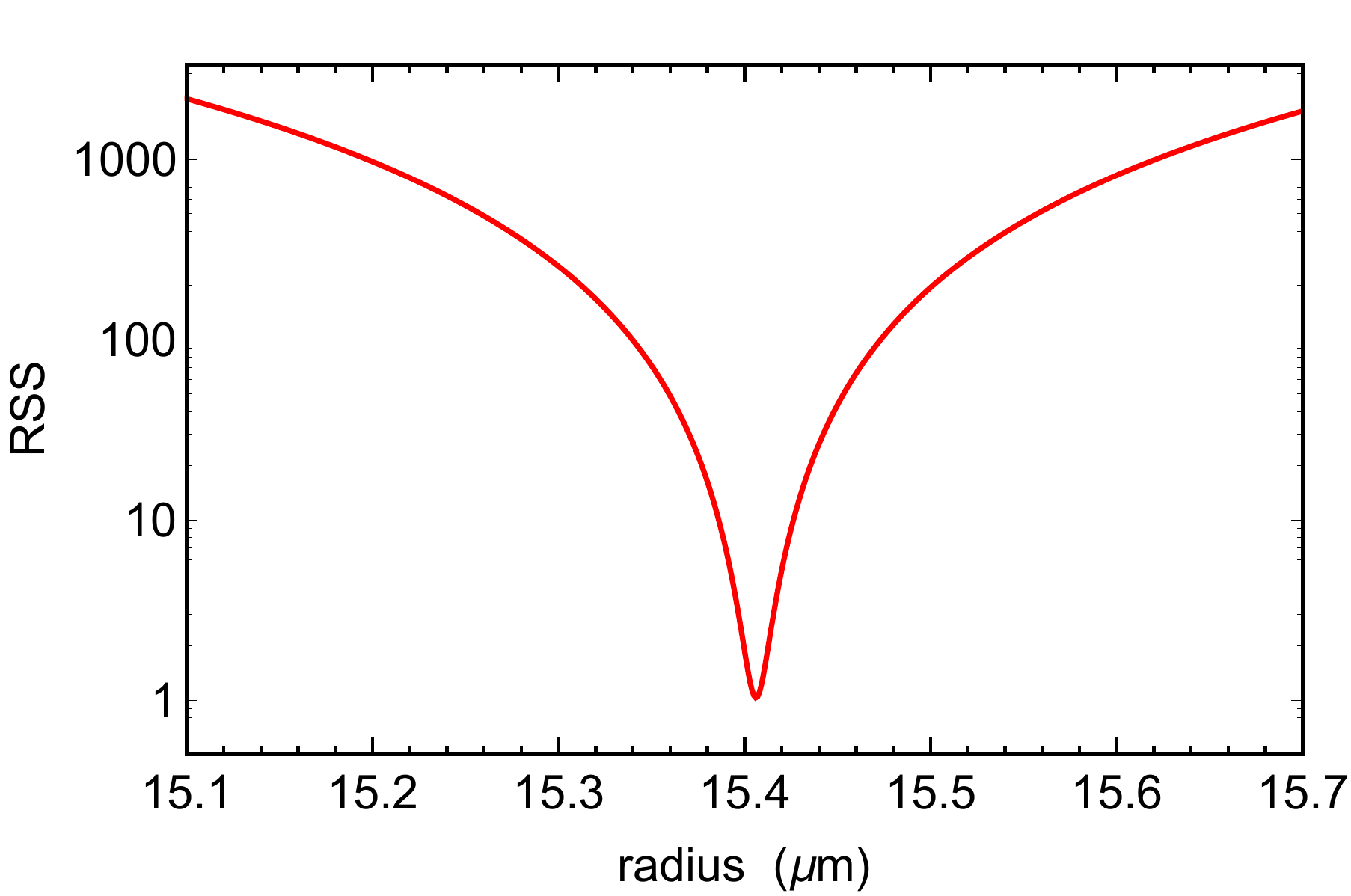}
\caption{ Minimum of the residual sum of squares (RSS) between the measured and calculated  azimuthal free spectral ranges  as  a  function  of the fiber radius $R$ for the optimal set of $m$ and $p$.}
\label{fig:RSS}
\end{figure}

In order to determine the exact quantum numbers  $m$  and $p$ of the analyzed WGM, we compared the azimuthal free spectral range (azFSR) measured in the experiment to the  azFSR calculated using the numerical solution of Eq.~(\ref{eq:transcendental}). In particular, we measured the azFSR of the fundamental axial mode for three different mode families. The calculated azFSR has two discrete and one continuous free parameters, $m$, $p$ and the radius $R$. 
Thus, we try to find the  best estimate for  $R$ by  calculating the  azFSRs for a large set of quantum numbers $m$ and $p$,  and  compute the  residual  sum of  squares  (RSS)  to  the measured  azFSRs.  The minimal RSS for the set of parameters given in Tab.~\ref{tab:quantumnumbers} as a function of the radius $R$ is shown in Fig.~\ref{fig:RSS}. Mode pair \# 1 belongs to the same mode family as the modes used in the manuscript, thus yielding the quantum numbers $m$ and $p$.

\begin{table}[htbp]
\centering
\caption{\bf Set of quantum numbers used to obtain a minimal RSS of the measured  and  calculated  azFSRs}
\begin{tabular}{cccc}
\\
\hline
mode pair \# & $m$&$m+1$ & $p$  \\ \hline
 1 & 117&118 &8  \\
 2 & 121&122 & 7 \\
 3 & 134&135 & 4  \\

\hline
\end{tabular}
  \label{tab:quantumnumbers}
\end{table}

\section{Estimation of the measurement error}In order to derive an expression for the error in the estimation of the axial radius variations, studied in the main manuscript, we start with Eq.~(3), which can be approximated, for the case of $\lambda\ll \lambda_z$, by
\begin{equation}
	R(z)=R_q +\frac{R_q}{2} \left(\frac{\lambda_q}{\lambda_z'(z)\,n_0}\right)^2 \;,\label{eq:R_approx}
\end{equation}
where $R_q=f(m,p) m\lambda_q/2\pi n_0$ is the radius at the caustic of the axial mode with quantum number $q$ and vacuum wavelength $\lambda_q$. The refractive index of the resonator material, $n_0$, enters here because the axial wavelength measured via imaging corresponds to $\lambda_z'=\lambda_z/n_0$. In this form, one can separate the error of the local radius $R(z)$ into an error of the absolute caustic radius $R_q$ and an error for the axial variations around $R_q$ given by the second term in Eq.~(\ref{eq:R_approx}). 
The relative error of $R_q$ is given by
\begin{align}
\frac{\Delta R_q}{R_q}=&\left(\frac{1}{f}\frac{\partial f}{\partial m}+\frac{1}{m}\right)\Delta m+\frac{1}{f}\frac{\partial f}{\partial p}\Delta p \nonumber\\
&+\left(\frac{1}{f}\frac{\partial f}{\partial n_0}+\frac{1}{n_0}\right)\Delta n_0+\frac{\Delta \lambda_q}{\lambda_q}
\end{align}
and the error for the radius variations $R-R_q$ is given by
\begin{equation}
\frac{\Delta (R-R_q)}{R-R_q}=\left(1+\frac{\Delta R_q}{R_q}+2\frac{\Delta n_0}{n_0}+2\frac{\Delta \lambda_q}{\lambda_q}\right)\cdot2\frac{\Delta\lambda_z'}{\lambda_z'}\;, \label{eq:error_variations}
\end{equation}
where the errors on $R_q$, $\lambda_q$ and $n_0$ only enter as a global factor into the error of the radius variation, as these quantities are constant for the whole measurement. As a consequence, the error for the radius profile is in most relevant cases dominated by the error in the measurement of $\lambda_z$.\\

\begin{table}[htbp]
	\centering
	\caption{\bf  Error budget of the absolute radius $\sigma_R $ and the radius variation $\sigma_{ R-R_q}$, for $\lambda_z'=50$~$\boldsymbol{\mu}$m}
	\begin{tabular}{cccc}
	\\
		\hline
		Error source & error & $\sigma_{R_q} $ (pm)&  $\sigma_{ R-R_q} $ (pm) \\
		\hline
		$n_0$ 				& $10^{-5}$ 						& 106				&$10^{-3}$\\
		$\lambda_q$ 	& $1.4\times10^{-5}$~nm & 0.3				&$10^{-6}$\\
		$\lambda_z'$ & 0.2~$\mu$m 						& 8.4 			&0.5\\
		$f$				 	& $<10^{-7} $			& &\\ \hline
		
	\end{tabular}
	\label{tab:error}
\end{table}

Table \ref{tab:error} shows the uncertainties of the different quantities for our experiment and their contribution to the error in measuring the caustic radius $R_q$ and the axial radius variations. In more detail, the light's wavelength $\lambda_q$ was measured using a wavemeter (High Finesse WS7-60) with a specified absolute accuracy of 60~MHz which for our wavelength corresponds to $\Delta\lambda_q=1.4\times10^{-5}$~nm. The discrete quantum numbers $m$ and $q$ are directly obtained from our fitting procedure and, thus, do not contribute to the error. Another error source originates from the not precisely known refractive index $n_0$ which we obtained from the Sellmeier equations given in Ref.~\cite{Malitson1965}. As an estimate for the error of the refractive index of our fiber we use the maximal residuals in Ref.~\cite{Malitson1965}, that amounts to  $\Delta n_0=10^{-5}$ at 850 nm wavelength. Compared to this, the error of $n_0$ due to the wavelength error can be neglected. 
The geometric dispersion also depends on the refractive index of the fiber and the surrounding material. Using the asymptotic expansion given by Eq.~\ref{eq:Lam}, this can be estimated to be $|df/dn_0|<0.01$ for a large range of quantum numbers $m$ and $p$. When we use the refractive index of air instead of vacuum, $f$ changes by $10^{-7}$. To determine the axial wavelength $\lambda_z$, we locally fitted a parabolic curve to the axial intensity profile. To estimate the error of this fitting procedure, Poisson noise according to our experimental signal to noise level was added to the measured data and then fitted again. The standard deviation of the resulting minimum positions then yields an uncertainty in the axial wavelength determination of $\Delta\lambda_z=0.2~\mu$m.
Consequently, in our experiment, the uncertainty in $R_q$ is dominated by the uncertainty of the refractive index $n_0$, while for the radius variation the dominant error source is the measurement error of the axial wavelength, $\lambda_z$, when imaging the wavefunctions.\\

As mentioned in the main manuscript, the optical measurement of the radius profile is very robust even if the the exact experimental parameters are not known. To illustrate this, we also consider the case, where one does not have access to a wavemeter, so the resonance wavelength can not be determined precisely. In this case, the individual errors are significantly larger. We assume the errors $\Delta n_0=0.01$ and $\Delta \lambda_q=1$~nm  which can, e.g., be obtained by using a simple grating spectrometer. In this case, the procedure to determine the radial and azimuthal quantum numbers, $p$ and $m$, introduced before will not work. This will result in an increased uncertainty on these parameters and consequently on the geometric dispersion factor $f(m,p)$. However, one can directly obtain an estimate of the absolute radius of the fiber $R_q$ using standard microscopy. If we assume an error of about $\Delta R_q=0.5$ $\mu$m for a fiber radius of 15 $\mu$m as in our experiment, the obtainable precision in the radius variations $R-R_q$ based on our slow-light imaging method is not significantly reduced, as the first term on the right-hand side of Eq.~(\ref{eq:error_variations}) is still approximately equal to one. Remarkably, even in this case, one retains sub-\AA ngstr\"om precision.

\section{Mean standard deviation for a parabolic radius profile}

In order to estimate the scaling of the mean standard error as a function of the axial quantum number $q$, that is plotted in Fig.~4b, we assume a parabolic radius profile
\begin{equation}
R(z) =R_0\left(1-\frac{(\Delta k z)^2}{2}\right)\;.
\end{equation}
Here, $R_0$ is the central radius and $\Delta k$ the curvature of the profile. From a fit to the experimental data we obtain $\Delta k= 0.000052$ $\mu$m$^{-1}$. In order to get the mean spacing between two intensity minima, we assume that the mode extends between the classical turning points, i.e., the two caustics and divide it by the number of intensity maxima, $q+1$. The resulting mean axial wavelength $\overline{\lambda_z}$ can be inserted into Eq.~(5) to obtain the theoretically expected mean standard deviation shown in Fig.~4b.

\bibliography{bib}